\documentclass[12pt]{article}

\usepackage{amsmath}
\usepackage{amssymb}
\usepackage{graphicx,color}
\usepackage{here}

\title{A new randomness test solving problems of Discrete Fourier Transform Test}

\author{Atsushi Iwasaki\thanks{Fukuoka Institute of Technology, a-iwasaki@fit.ac.jp} \and Ken Umeno\thanks{Kyoto University, umeno.ken.8z@kyoto-u.ac.jp}}

\begin{document}

\maketitle

\begin{abstract}
Discrete Fourier Transform Test (DFTT), which is a randomness test included in NIST SP800-22, has a problem.
It is that theoretical reference distribution of the test statistic has not been derived.
In this paper, we propose a new test using variance of power spectrum as the test  statistic, whose reference distribution can be theoretically derived.
The purpose of DFTT is to detect periodic features and that of the proposed test is the same.
We make some experiments and show that the proposed test has stronger detection power than DFTT.
\end{abstract}

\section{Introduction}
Randomness test is one kind of statistical hypothesis test under the null hypothesis that the given sequence is truly random sequence.
It is applicable to any sequence irrespective of generator and  useful for many fields.
 In particular, it is indispensable for estimation of security of cryptography.

NIST SP800-22 \cite{NIST} is one of the most famous randomness test suite in the world.
The first version of SP800-22 was published in 2001 and used in the selection of  Advanced Encryption Standard \cite{AES}. 
Now, revision 1a which is the recent version of SP800-22 have been published.
SP800-22 have been pointed that it has problems and some improvements have been applied.
Some problems, however, have been left even in revision 1a.

Discrete Fourier Transform Test (DFTT), which is also called Spectral Test, is a test included in SP800-22.
The purpose of this test is to detect periodic features. 
The algorithm of DFTT is as follows:
\begin{enumerate}
\item Receive $M$ sequences $X_1,X_2,\cdots, X_M$ as the input.
Here, each $X_i$ is a $n$-bit sequence, and each bit is 0 or 1.
\item Convert each bit $x$ to $2x-1$. (Then, each bit becomes 1 or -1.)
\item For each $X_i$, compute p-value $p_i$.
\item Make a hypothesis test for $\{p_1,p_2,\cdots,p_M\}$ under the null hypothesis ``$p_1,p_2,\cdots,p_M$ are independent and follow the uniform distribution on the interval $[0,1]$ " (Second-level-test).
\end{enumerate}
The null hypothesis of DFTT is rejected if and only if the null hypothesis of step 4 is rejected.
It means that if the null hypothesis of DFTT is true, $p_1,p_2,\cdots,p_M$ must be independent and follow the uniform distribution on the interval $[0,1]$.
The independency of $p_1,p_2,\cdots,p_M$ is clearly true if $X_1,X_2,\cdots,X_M$ are independent.
Then, the most important point of DFTT is computation of p-value at step 3.
If $X_i$ is a truly random sequence, the corresponding p-value $p_i$ must follow the uniform distribution on $[0,1]$.

The algorithm of computing p-value in the first version is as follows:
\begin{enumerate}
\item For given $n$-bit sequence $X$, perform discrete Fourier Transform and get the Fourier spectrum series $|S_0(X)|,|S_1(X)|,\cdots,|S_{\frac{n}{2}-1}(X)|$.
\item Count $N_1$ which is the number of $|S_i(X)|<\sqrt{3n}$.
\item Compute $d$ as follows:
\begin{align*}
d=\frac{N_1-0.95\frac{n}{2}}{\sqrt{(0.95)(0.05)\frac{n}{2}}}.
\end{align*}
\item Compute p-value $p$ as follows:
\begin{align*}
p=\text{erfc}\left(\frac{|d|}{\sqrt{2}}\right),
\end{align*}
where erfc means the complementary error function.
\end{enumerate}
This algorithm is based on the following conjecture:
for sufficient large $n$, if $X$ is truly random sequence, $\frac{2}{n}|S_1(X)|^2,\cdots,\frac{2}{n}|S_{\frac{n}{2}-1}(X)|^2$ independently follow　$\chi^2_2$-distribution and  $N_1$ follows $\mathcal{B}(\frac{n}{2},0.95)$, where $\mathcal{B}$ means binomial distribution.
If the conjecture is true, $d$ and $p$ approximately follow the standard normal distribution and uniform distribution on $[0,1]$ respectively.
The conjecture is, however, probably not true because even sequences generated by Mersenne twister \cite{MT} which can be regarded as good generator do not pass the DFTT.
There are following two problems:
\begin{itemize}
\item For an arbitrary fixed $j<n$ except $j=0$, it is true that $\frac{2}{n}|S_j(X)|^2$ follows $\chi^2_2$-distribution if $X$ is truly random sequence and $n$ is sufficient large.
However, $|S_0(X)|,|S_1(X)|,\cdots,$
$|S_{\frac{n}{2}-1}(X)|$ are not probably independent because they share the same argument $X$.
\item The threshold $\sqrt{3n}$ is an approximate value, but the precision is not good.
\end{itemize}

In 2003, Kim et al. pointed out the above problems \cite{Kim}.
Hamano also pointed out them \cite{Hamano}.
Kim et al. proposed the following improvement:
\begin{itemize}
\item Change the computation of $d$ as follows:
\begin{align*}
d=\frac{N_1-0.95\frac{n}{2}}{\sqrt{(0.95)(0.05)\frac{n}{4}}}.
\end{align*}
\item Set the threshold as $\sqrt{-n\log(0.05)}$.
\end{itemize}
The first improving point is based on fitting using pseudo random sequences.
The second point has a theoretical basis if the first point is proper.
DFTT in the recent version of NIST SP800-22 adopted the improvement.

Pareschi et al. did more precise fitting using pseudo random sequences generated by BBS \cite{BBS} and propose further improvement \cite{Pareschi}.
That is to change  the computation of $d$ as follows:
\begin{align*}
d=\frac{N_1-0.95\frac{n}{2}}{\sqrt{(0.95)(0.05)\frac{n}{3.8}}}.
\end{align*}

Although DFTT has been improved, the authors think that the following problems are left:
\begin{itemize}
\item The purpose of randomness test is to estimate randomness of given sequences.
Then, fitting using pseudo random sequences is inconsistent even if it is theoretically ensured that PRNG generating the sequences is superior.
\item Even if the PRNG is perfectly  superior, fitting is merely numerical.
\item It is not ensured that $N_1$ follows binomial distribution under the null hypothesis. Then, there is no basis that the computation of $d$ is written as the form
\begin{align*}
d=\frac{N_1-0.95\frac{n}{2}}{\sqrt{(0.95)(0.05)\frac{n}{a}}},
\end{align*}
where $a$ is a parameter.
\end{itemize}

Okada et al. proposed an approach to solve the problems \cite{Okada}.
The authors, however, think that the approach made another problem.
That is to break independency of p-values.
In this paper we propose another randomness test which detects periodic features, whose reference distribution can be theoretically derived.

This paper is constructed as follows:
In Section 2, we deal with variance of power spectrum and derive its distribution.
In Section 3, we propose a new randomness test using the  variance of power spectrum.
In Section 4, we make experiments to estimate the detection power of the proposed test.
Finally, we conclude this paper.

\section{Variance of power spectrum}
The reason why the problems are left is that it is too difficult to derive the reference distribution of $N_1$.
Counting the number of Fourier coefficients over a threshold is a non-linear operation and the non-linearity causes the difficulty.
The purpose of DFTT is to investigate whether the fluctuation of Fourier spectrum is suitable or not.
Then, we should introduce another indicator which reflects the fluctuation and whose reference distribution can be analytically derived.

Firstly, we consider the variance of Fourier spectrum as the indicator.
It is, however, difficult to derive the reference distribution of the variance because the definition of Fourier spectrum is 
\begin{align*}
&|S_j(X)|:=\sqrt{\left(\sum_{k=0}^{n-1}x_k\cos\left(\frac{2\pi kj}{n}\right)\right)^2+\left(\sum_{k=0}^{n-1}x_k\sin\left(\frac{2\pi kj}{n}\right)\right)^2},
\end{align*}
where $x_k$ is $(k+1)$-th bit of $X$ and the square root makes it difficult.

Then, we consider the variance of power spectrum as the indicator.
The definition is as follows:
\begin{align*}
V_n(X):=\frac{1}{n}\sum_{j=0}^{n-1}\left\{\frac{1}{n}|S_j(X)|^2-\frac{1}{n}\sum_{r=0}^{n-1}\left(\frac{1}{n}|S_r(X)|^2\right)\right\}^2.
\end{align*}
Of course, the variance of power spectrum is not the variance of Fourier spectrum, but the variance of power spectrum probably reflects fluctuation of Fourier spectrum.
\subsection{Distribution of $V_n(X)$}
In order to derive the distribution of $V_n(X)$, let us compute the average and variance of $V_n(X)$.
Firstly, we deform $V_n(X)$.
\begin{align*}
&V_n(X)\\
:=&\frac{1}{n}\sum_{j=0}^{n-1}\left\{\frac{1}{n}|S_j(X)|^2-\frac{1}{n}\sum_{r=0}^{n-1}\left(\frac{1}{n}|S_r(X)|^2\right)\right\}^2\\
=&\frac{1}{n^3}\sum_{j=0}^{n-1}|S_j(X)|^4-1\ \ \ \ (\because\text{Parseval's theorem})\\
=&\frac{1}{n^3}\sum_{j=0}^{n-1}\Big\{\left(\sum_{k=0}^{n-1}x_k\cos\left(\frac{2\pi kj}{n}\right)\right)^2+\left(\sum_{k=0}^{n-1}x_k\sin\left(\frac{2\pi kj}{n}\right)\right)^2\Big\}^2-1\\
=&\frac{1}{n^3}\sum_{j=0}^{n-1}\sum_{a=0}^{n-1}\sum_{b=0}^{n-1}\sum_{c=0}^{n-1}\sum_{d=0}^{n-1}x_ax_bx_cx_d\Big\{ \\
&\cos\left(\frac{2\pi aj}{n}\right)\cos\left(\frac{2\pi bj}{n}\right)\cos\left(\frac{2\pi cj}{n}\right)\cos\left(\frac{2\pi dj}{n}\right)\\
&+\sin\left(\frac{2\pi aj}{n}\right)\sin\left(\frac{2\pi bj}{n}\right)\sin\left(\frac{2\pi cj}{n}\right)\sin\left(\frac{2\pi dj}{n}\right)\\
&+\cos\left(\frac{2\pi aj}{n}\right)\cos\left(\frac{2\pi bj}{n}\right)\sin\left(\frac{2\pi cj}{n}\right)\sin\left(\frac{2\pi dj}{n}\right)\Big\}-1
\end{align*}
\begin{align*}
=&\frac{1}{4n^3}\sum_{j=0}^{n-1}\sum_{a=0}^{n-1}\sum_{b=0}^{n-1}\sum_{c=0}^{n-1}\sum_{d=0}^{n-1}x_ax_bx_cx_d\Big\{\\
&\cos\left(\frac{2\pi (a+b+c-d)j}{n}\right)+\cos\left(\frac{2\pi (a+b-c+d)j}{n}\right)\\
& -\cos\left(\frac{2\pi (a-b+c+d)j}{n}\right)-\cos\left(\frac{2\pi (a-b-c-d)j}{n}\right)\\
&+2\cos\left(\frac{2\pi (a-b+c-d)j}{n}\right)+2\cos\left(\frac{2\pi (a-b-c+d)j}{n}\right)\Big\}-1.
\end{align*}
We introduce a new notation:
\begin{align*}
\delta_x:=
\begin{cases}
1\ \ \ (x=0,\pm n)\\
0\ \ \ (\text{otherwise})
\end{cases}
.
\end{align*}
Then,
\begin{align}
V_n(X)\nonumber
=&\frac{1}{4n^2}\sum_{a=0}^{n-1}\sum_{b=0}^{n-1}\sum_{c=0}^{n-1}\sum_{d=0}^{n-1}x_ax_bx_cx_d\Big\{\nonumber\\
&\ \ \ \ \delta_{a+b+c-d}+\delta_{a+b-c+d}
-\delta_{a-b+c+d}\nonumber\\&-\delta_{a-b-c-d}
+2\delta_{a-b+c-d}+2\delta_{a-b-c+d}\Big\}-1\nonumber\\
=&\frac{1}{n^2}\sum_{a=0}^{n-1}\sum_{b=0}^{n-1}\sum_{c=0}^{n-1}\sum_{d=0}^{n-1}x_ax_bx_cx_d\delta_{a-b+c-d}-1.\label{siki1}
\end{align}

Let us compute the average of $V_n(X)$.
Since each $x_i$ takes 1 or -1 with the same probability, terms in (\ref{siki1}) except terms hold $x_ax_bx_cx_d=1$ vanish when we take the average.
Then,
\begin{align*}
&\mathbb{E}[V_n(X)]\\
=&\frac{1}{n^2}\Big\{ \ \sum_{a=b=c=d}\delta_{a-b+c-d}+\sum_{a=b\ne c=d}\delta_{a-b+c-d}\\&\ \ \ \ \ \ \ \ \ \ \ \ \ \ +\sum_{a=c\ne b=d}\delta_{a-b+c-d}+\sum_{a=d\ne b=c}\delta_{a-b+c-d}\ \Big\}-1\\
=&
\begin{cases}
1\ \ \ \ &(n\text{: even})\\
1-\frac{1}{2n}\ \ \ &(n\text{: odd})
\end{cases}
\ \ \longrightarrow 1\ \ (n\to \infty).
\end{align*}

Next, let us compute the variance of $V_n(X)$.
We introduce new notations:
\begin{align*}
B_1:=&\left\{(a,b,c,d)\in\{0,1,\cdots,n-1\}^4|a,b,c,d\ \text{take different values each other}\right\},\\
B_2:=&\left\{(e,f,g)\in\{0,1,\cdots,n-1\}^3|e,f,g\ \text{take different values each other}\right\}.
\end{align*}
Then, 
\begin{align*}
&\mathbb{E}\left[\left(V_n(X)-\mathbb{E}[V_n(X)]\right)^2\right]\\
=&\mathbb{E}\Big[\big\{\sum_{(a,b,c,d)\in B_1}x_ax_bx_cx_d\delta_{a-b+c-d}+\sum_{(e,f,g)\in B_2}x_fx_g\delta_{2e-f-g}\big\}^2\Big]\\
=&\frac{1}{n^4}\Big\{\mathbb{E}[(\sum_{(a,b,c,d)\in B_1}x_ax_bx_cx_d\delta_{a-b+c-d})^2] +\mathbb{E}[(\sum_{(e,f,g)\in B_2}x_fx_g\delta_{2e-f-g})^2]\\&\ \ \ \ \ +\mathbb{E}[(\sum_{(a,b,c,d)\in B_1}x_ax_bx_cx_d\delta_{a-b+c-d})(\sum_{(e,f,g)\in B_2}x_fx_g\delta_{2e-f-g})]\Big\}.
\end{align*}
Let us consider the first term.
\begin{align*}
&\mathbb{E}[(\sum_{(a,b,c,d)\in B_1}x_ax_bx_cx_d\delta_{a-b+c-d})^2]\\
=&\mathbb{E}[\sum_{(a,b,c,d)\in B_1}\sum_{(a^\prime,b^\prime,c^\prime,d^\prime)\in B_1}x_ax_bx_cx_dx_{a^\prime}x_{b^\prime}x_{c^\prime}x_{d^\prime}\delta_{a-b+c-d}\delta_{a^\prime-b^\prime+c^\prime-d^\prime}]\\
=&\mathbb{E}[\sum_{(a,b,c,d)\in B_1}\sum_{(a^\prime,b^\prime,c^\prime,d^\prime)\in R(a,b,c,d)}\delta_{a-b+c-d}\delta_{a^\prime-b^\prime+c^\prime-d^\prime}],
\end{align*}
where $R(a,b,c,d)=\{\text{permutation of }a,b,c,d\}$.
\begin{table}
\begin{center}
\caption{Combinatorial number satisfying $(a,b,c,d)\in B_1$, $(a^\prime,b^\prime,c^\prime,d^\prime)\in R(a,b,c,d)$}
\label{A}
\begin{tabular}[t]{|c|c|c|}
\hline
$a-b+c-d$ & $a^\prime-b^\prime+c^\prime-d^\prime$ &Combinatorial number\\
\hline\hline
0 &0 & $\frac{8}{3}n^3+O(n^2)$\\
\hline
0& $n$& $O(n^2)$\\
\hline
0& $-n$& $O(n^2)$\\
\hline
$n$ &0 & $O(n^2)$\\
\hline
$n$& $n$& $\frac{4}{3}n^3+O(n^2)$\\
\hline
$n$& $-n$& $\frac{4}{3}n^3+O(n^2)$\\
\hline
-$n$ &0 & $O(n^2)$\\
\hline
-$n$& $n$& $\frac{4}{3}n^3+O(n^2)$\\
\hline
-$n$& $-n$& $\frac{4}{3}n^3+O(n^2)$\\
\hline
\end{tabular}
\end{center}
\end{table}
By Table. \ref{A},
\begin{align*}
&\mathbb{E}[(\sum_{(a,b,c,d)\in B_1}x_ax_bx_cx_d\delta_{a-b+c-d})^2]=8n^3+O(n^2).
\end{align*}
By the same way, the second term is $O(n^2)$ and the third term equals to 0 .
Then,
\begin{align*}
\mathbb{E}\left[\left(V_n(X)-\mathbb{E}[V_n(X)]\right)^2\right]
=\frac{8}{n}+O\left(\frac{1}{n^2}\right)\ \ \ \ \longrightarrow 0\ (n\to\infty).
\end{align*}

\subsection{Scaling}
The variance of $V_n(X)$ vanishes as $n\to\infty$, and so we cannot, unfortunately, use $V_n(X)$ as a new indicator.
Then, we consider the following scaled variance of power spectrum as the indicator:
\begin{align*}
\tilde{V}_n(X):=&\sqrt{\frac{n}{8}}\left\{V_n(X)-\mathbb{E}[V_n(X)]\right\}\\
=&\frac{1}{(2n)^{\frac{3}{2}}}\Big\{\sum_{(a,b,c,d)\in B_1}x_ax_bx_cx_d\delta_{a-b+c-d}+\sum_{(e,f,g)\in B_2}x_fx_g\delta_{2e-f-g}\Big\}.
\end{align*}
In order to derive the distribution of $\tilde{V}_n(X)$, let us compute the moment.
\begin{align*}
&\mathbb{E}[(\tilde{V}_n)^m]\\
=&\frac{1}{(2n)^{\frac{3}{2}m}}\sum_{l=0}^{m} {}_mC_l
\mathbb{E}\Big[\Big\{\sum_{(a,b,c,d)\in B_1}x_ax_bx_cx_d\delta_{a-b+c-d}\Big\}^{m-l}\\
&\ \ \ \ \ \ \ \ \ \times\Big\{\sum_{(e,f,g)\in B_2}x_fx_g\delta_{2e-f-g}\Big\}^l\Big]\\
=&\frac{1}{(2n)^{\frac{3}{2}m}}\mathbb{E}\Big[\Big\{\sum_{(a,b,c,d)\in B_1}x_ax_bx_cx_d\delta_{a-b+c-d}\Big\}^{m}\Big]+(\text{Lower order terms}),
\end{align*}
\begin{align*}
&\mathbb{E}\Big[\Big\{\sum_{(a,b,c,d)\in B_1}x_ax_bx_cx_d\delta_{a-b+c-d}\Big\}^{m}\Big]\\
=&\sum_{(a(1),b(1),c(1),d(1))\in B_1}\cdots\sum_{(a(m),b(m),c(m),d(m))\in B_1}\\
&\ \ \ \ \ \ \ \ \ \ \mathbb{E}\left[\{x_{a(1)}x_{b(1)}x_{c(1)}x_{d(1)}\}\cdots \{x_{a(m)}x_{b(m)}x_{c(m)}x_{d(m)}\}\right]\\
&\ \ \ \ \ \ \ \ \ \ \ \ \ \ \ \ \times \delta_{a(1)-b(1)+c(1)-d(1)}\cdots\delta_{a(m)-b(m)+c(m)-d(m)}.
\end{align*}
Then, we should count number of $\{a(1),b,(1),c(1),d(1)\},\cdots,$\\
$\{a(m),b(m),c(m),d(m)\}$ satisfying
\begin{align*}
&\mathbb{E}\left[\{x_{a(1)}x_{b(1)}x_{c(1)}x_{d(1)}\}\cdots \{x_{a(m)}x_{b(m)}x_{c(m)}x_{d(m)}\}\right]=1,\\
&\delta_{a(1)-b(1)+c(1)-d(1)}\cdots\delta_{a(m)-b(m)+c(m)-d(m)}=1.
\end{align*}
In order to count the number, we consider the following model:
\begin{itemize}
\item Basically, each variable can freely take $n$ values. (In other words, each variable has one degree of freedom.)
\item Each variable has a hand.
\item Variables which have a same argument are included in a same set. (See Fig. \ref{B1}.)
\item Each hand must be connected with another hand by final time.
\item Each hand must not be connected with two or more other hands.
\item Hands of variables included in a same set must not be connected each other.
\item There are the following constraint conditions:
\begin{itemize}
\item Each set has one constraint condition. If values of three variables in a set are fixed, the other variable in the set is automatically determined.
\item If a hand of variable A is connected with another hand of variable B, the value of A must equal to the value of B.
\end{itemize}
\end{itemize}
\begin{figure}
\begin{center}
\begin{picture}(80,20)
\put(0,9){\large $\{a(i),b(i),c(i),d(i)\}$}
\put(15,7){\line(0,-1){7}}
\put(45,7){\line(0,-1){7}}
\put(75,7){\line(0,-1){7}}
\put(105,7){\line(0,-1){7}}
\end{picture}
\end{center}
\caption{Model of set of variables}
\label{B1}
\end{figure}
As example,  variables  in Fig. \ref{B1} have 3(=4-1) degree of freedom.
\begin{figure}
\begin{center}
\begin{picture}(80,65)
\put(0,56){\large $\{a(i),b(i),c(i),d(i)\}$}
\put(15,54){\line(0,-1){7}}
\put(45,54){\line(0,-1){7}}
\put(75,54){\line(0,-1){7}}
\put(105,54){\line(0,-1){7}}
\put(0,0){\large $\{a(j),b(j),c(j),d(j)\}$}
\put(15,10){\line(0,1){7}}
\put(45,10){\line(0,1){7}}
\put(75,10){\line(0,1){7}}
\put(105,10){\line(0,1){7}}
\put(45,17){\line(-1,1){30}}
\end{picture}
\end{center}
\caption{An example connecting a set with another set by one-connection}
\label{B2}
\end{figure}
We consider the case that add a set $\{a(j),b(j),c(j),d(j)\}$ to Fig. \ref{B1} and connect $a(i)$'s hand with $b(j)$'s hand. (See Fig. \ref{B2}.)
In this case, the value of $b(j)$ must equal to $a(i)$.
Then, degree of freedom increases and the amount of the change is +2(=4-1-1).
Hands which are not connected with other hands (we call ``open hands") also increase and the amount of the change is +2.

\begin{figure}
\begin{center}
\begin{picture}(80,65)
\put(0,56){\large $\{a(i),b(i),c(i),d(i)\}$}
\put(15,54){\line(0,-1){7}}
\put(45,54){\line(0,-1){7}}
\put(75,54){\line(0,-1){7}}
\put(105,54){\line(0,-1){7}}
\put(0,0){\large $\{a(j),b(j),c(j),d(j)\}$}
\put(15,10){\line(0,1){7}}
\put(45,10){\line(0,1){7}}
\put(75,10){\line(0,1){7}}
\put(105,10){\line(0,1){7}}
\put(45,17){\line(-1,1){30}}
\put(15,17){\line(1,1){30}}
\put(105,17){\line(0,1){30}}
\put(75,17){\line(0,1){30}}
\end{picture}
\end{center}
\caption{An example of optimum connection}
\label{B4}
\end{figure}
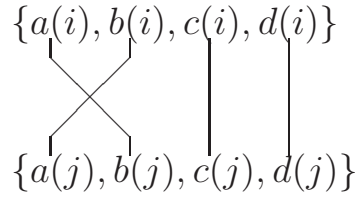

We consider a general case.
Table \ref{B3} shows the relation between number of anew connecting hands, amount of change of degree of freedom and open hands.
In initial states, there is a set and variables included in the set have 3 degree of freedom.  Fig. \ref{B1} is an example of initial state.
\begin{table}
\caption{Relation between number of connected hands and change of degree of freedom and open hands}
\label{B3}
\begin{center}
\begin{tabular}[t]{|c|c|c|}
\hline
Number of connected hands & Degree of freedom& Open hands\\
\hline\hline
1&+2&+2\\
\hline
2&+1&$\pm0$\\
\hline
3&$\pm0$&-2\\
\hline
4&$\pm0$&-4\\
\hline
\end{tabular}
\end{center}
\end{table}
Since we must not leave hands which are not connected with other hands, connecting a set with another set like Fig. \ref{B4} maximizes degree of freedom per one variable.

From the above, when $m$ is even,
\begin{align*}
&\mathbb{E}[(\tilde{V}_n)^m]\\
=&\frac{1}{(2n)^{\frac{3}{2}m}}\mathbb{E}\Big[\Big\{\sum_{(a,b,c,d)\in B_1}x_ax_bx_cx_d\delta_{a-b+c-d}\Big\}^{m}\Big]+(\text{Lower order terms})\\
=&\frac{1}{(2n)^{\frac{3}{2}m}}\mathbb{E}\Big[\Big\{\sum_{(a,b,c,d)\in B_1}x_ax_bx_cx_d\delta_{a-b+c-d}\Big\}^{2}\Big]^{\frac{m}{2}}\\
&\ \ \ \times (\text{The combinatorial total number to make $m/2$ pairs from $m$ items.})\\
&\ \ \ \ \ \ \ \ +(\text{Lower order terms})\\
=&(m-1)!!+(\text{Lower order terms})\ \ \longrightarrow(m-1)!!\ \ (n\to\infty).
\end{align*} 
Here, $x!!:=x\times(x-2)\times(x-4)\times\cdots\times3\times1$.
When $m$ is odd, since we cannot make $m/2$ pairs from $m$ items,
\begin{align*}
\mathbb{E}[(\tilde{V}_n)^m]
=0+(\text{Lower order terms})\ \ \ \ \longrightarrow0\ \ (n\to\infty).
\end{align*}
Summarizing the above,
\begin{align*}
\lim_{n\to\infty}\mathbb{E}[(\tilde{V}_n)^m]
=
\begin{cases}
(m-1)!! &(m\text{: even})\\
0 &(m\text{: odd})
\end{cases}
.
\end{align*}
Then, the moments of $\tilde{V}_n(X)$ converge to those of the standard normal distribution.
It means that distribution of $\tilde{V}_n(X)$ ``converges" to the standard normal distribution.

\subsection{Numerical simulations}
\begin{figure}[H]
\begin{center}
\includegraphics[width=10cm]{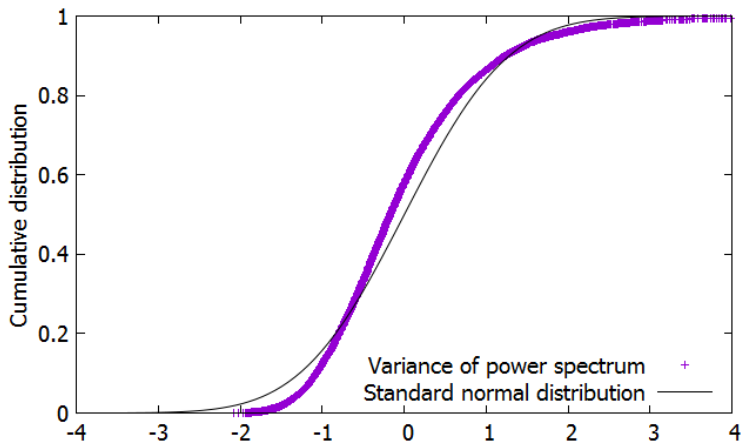}
\end{center}
\caption{Comparison cumulative distributions of $\tilde{V}_n(X)$ with the standard normal distribution. ($n=100$)}
\label{C1}
\end{figure}

\begin{figure}[H]
\begin{center}
\includegraphics[width=10cm]{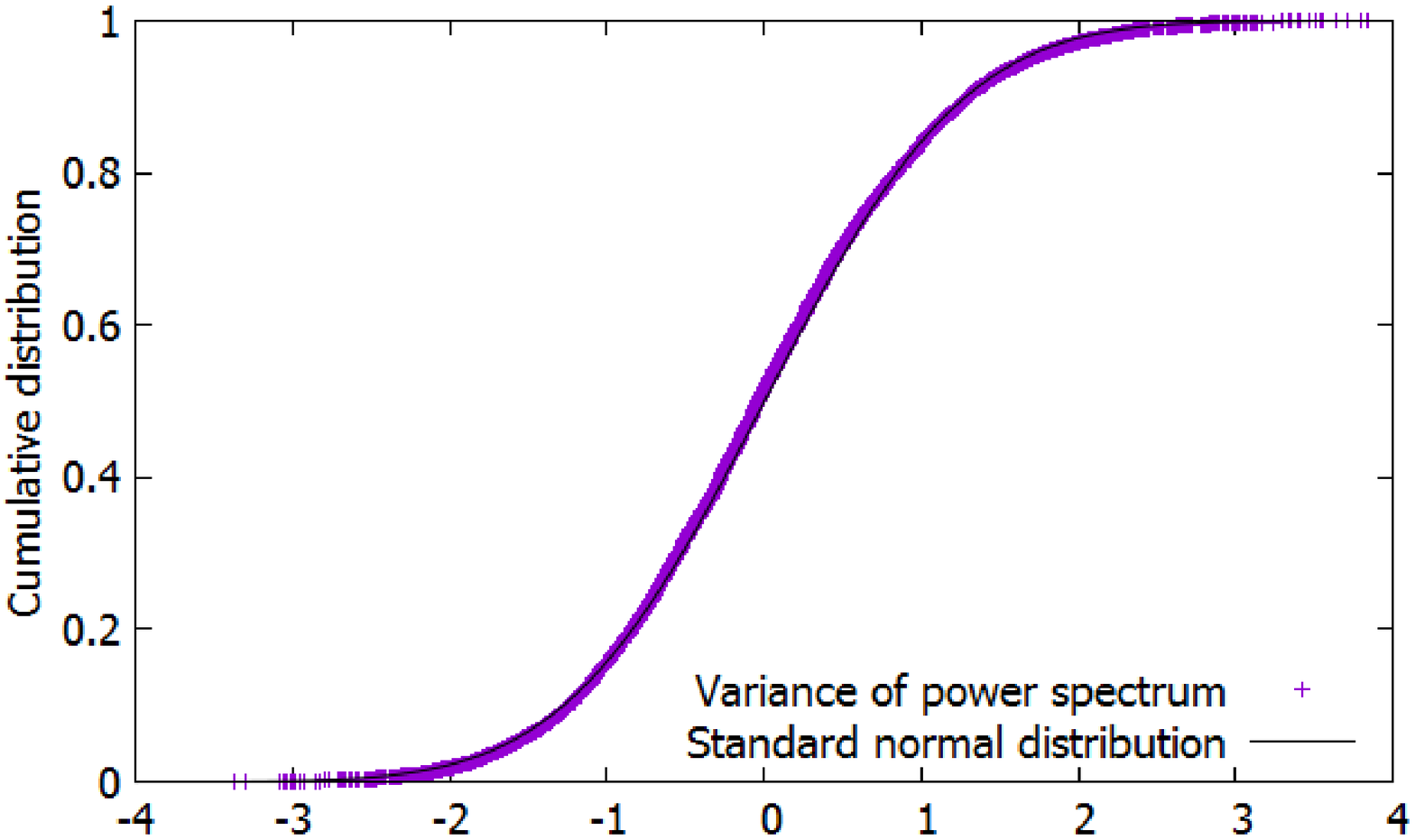}
\end{center}
\caption{Comparison cumulative distributions of $\tilde{V}_n(X)$ with the standard normal distribution. ($n=10000$)}
\label{C2}
\end{figure}

\begin{figure}[H]
\begin{center}
\includegraphics[width=10cm]{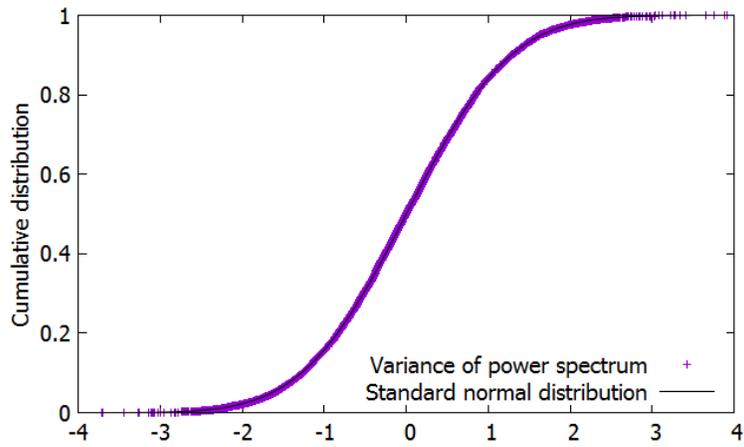}
\end{center}
\caption{Comparison cumulative distributions of $\tilde{V}_n(X)$ with the standard normal distribution. ($n=1000000$)}
\label{C3}
\end{figure}

We made cumulative distributions of $\tilde{V}_n(X)$ by 10000 $n$-bit sequences generated by Mersenne twister \cite{MT} and compared them and the standard normal distribution.
Fig. \ref{C1}, \ref{C2} and \ref{C3} show the results.
Roughly speaking, the cumulative distribution of $\tilde{V}_n(X)$ correspond to the standard normal distribution when $n\geq10^4$.

\section{Proposed test}
Based on the former section, we propose a new randomness test using $\tilde{V}_n(X)$.
The test replaces p-value computation of DFTT with the follows:
\begin{enumerate}
\item For given $n$-bit sequence $X$, perform discrete Fourier Transform and get the Fourier spectrum series $|S_0(X)|,|S_1(X)|,\cdots,|S_{\frac{n}{2}-1}(X)|$.
\item Compute $\tilde{V}_n(X)$ as follows:
\begin{align*}
\tilde{V}_n(X)=\frac{1}{\sqrt{2n^5}}\sum_{j=0}^{\frac{n}{2}-1}|S_j(X)|^4-\sqrt{\frac{n}{2}}.
\end{align*}
\item Compute p-value $p$ as follows:
\begin{align*}
p=\text{erfc}\left(\frac{|\tilde{V}_n(X)|}{\sqrt{2}}\right).
\end{align*}
\end{enumerate}
At the step 2, we use symmetry of Fourier spectrum series and deformed $\tilde{V}_n(X)$ in order to reduce computation.

The proposal method needs approximately $n$ multiplications more than computation of p-value of DFTT.
Discrete Fourier Transform, however, needs $O(n\log n)$ and so the proposal method needs $O(n\log n)$ times.
That is the same as DFTT.

\section{Evaluation of detection power of test}
In order to evaluate detection power of test, we made some experiments for the proposed test, the present DFTT (proposed by Kim et al.) and Pareschi et al.'s test.

Before explaining the experiments, we explain the second-level-test.
When we consider the case that $M$ $n$-bit sequences are tested, we get $M$ p-values $p_1,p_2,\cdots,p_M$.
The  second-level-test is a hypothesis test and the null hypothesis is that ``$p_1,p_2,\cdots,p_M$ are independent and follow the uniform distribution on the interval $[0,1]$".
For the $M$ p-values, we perform the following two tests:
\begin{itemize}
\item {\bf Proportion test}\\
Count $r$ which is the number of $p_i>0.01$.
Under the null hypothesis, $r$ follows $\mathcal{B}(0.99,M)$.
Then, if 
\begin{align*}
|r-0.99M|<3\times(\text{the standard deviation}),
\end{align*}
 we make the null hypothesis pass.
Else, we reject the null hypothesis.
\item {\bf Uniformity test}\\
Under the null hypothesis, we perform $\chi^2$-test for $\{p_1,p_2,\cdots,p_M\}$ with 10 bins and get new one p-value $p_{uniform}$.
If $p_{uniform}>0.0001$, we make the null hypothesis pass.
Else, we reject the null hypothesis.
\end{itemize}

\subsection{Experiment 1}

We investigate the pass rate for sequences generated by  Mersenne twister and AES.
The purpose of this experiment is to estimate Type 1 errors of the proposed test, the present DFTT and Pareschi et al.'s test.
Since a randomness test is a statistical test, probability that an ideal random sequence is rejected does not vanish.
In this experiment, we investigate Type 1 errors are in excusable range or not.

For each length and generator, $10^7$ sequences are used for the experiments.
 They are divided into 1,000 sets consist of 1,000 sequences.
For fair evaluation, the same sequences were used for evaluation of the proposed test, the present DFTT and Pareschi et al.'s test.

Table. \ref{MT1}-\ref{AES4} show the results.
Table \ref{MT2}-\ref{AES3} directly show the number of rejected sets and the other tables show subsidiary results.
By the results, we can say that the present DFTT is not a proper test.
There are some cases that results  for the long sequences ($n\geq10^6$) of Pareschi et al.'s test and the proposed test are out of the standard deviation range. These results, however, do not mean that they are not proper tests because the deviations from the standard deviation range are small and the probabilities of arising such cases are not negligible even if the tests are perfectly proper.

Table. \ref{MT3}-\ref{AES4} show that p-value's distribution of the proposed test quickly converges to [0,1] uniform distribution as sequence length becomes large.
That is a good property.

\begin{table}[H]
\begin{center}
\caption{Number of sequences whose p-values are smaller than 0.01. Sequences were generated by Mersenne twister.  Boldface values mean out of the standard deviation range.}
\label{MT1}
\begin{tabular}{lccc}
\hline
Length &Kim& Pareschi & Proposed\\
\hline
\hline
$10^3$&\bf 16030&\bf 7016&\bf 12702\\
$5\times10^3$&\bf 13553&\bf 9424&\bf 10797\\
$10^4$&\bf 10974&\bf 10974&\bf 10283\\
$5\times10^4$&\bf 12457& 9915& 9913\\
$10^5$&\bf 12291&\bf 9592& 10054\\
$5\times10^5$&\bf 12206&\bf 10154&\bf 10142\\
$10^6$&\bf 12306&\bf 10349& 10000\\
$5\times10^6$&\bf 12143& 10098& 9980\\
$10^7$&\bf 11977& 9907&\bf 10173\\
\hline
\end{tabular}
\end{center}
\end{table}
\begin{table}[H]
\begin{center}
\caption{Number of sequences whose p-values are smaller than 0.01. Sequences were generated by AES.   Boldface values mean out of the standard deviation range.}
\label{AES1}
\begin{tabular}{lccc}
\hline
Length &Kim& Pareschi & Proposed\\
\hline
\hline
$10^3$&\bf 15749&\bf  6942&\bf  12661 \\
$5\times10^3$&\bf 13599 &\bf 9530&\bf  10805\\
$10^4$&\bf 10720&\bf  10720&\bf  10282\\
$5\times10^4$&\bf 12543& 9945&\bf  10194\\
$10^5$&\bf 12397&\bf  9710&\bf  10131\\
$5\times10^5$&\bf 12156 &\bf 10154& 10081\\
$10^6$&\bf 12424&\bf  10466& 10006\\
$5\times10^6$&\bf 12323&\bf 10246&\bf 10127 \\
$10^7$&\bf  11961& 9959& 9905\\
\hline
\end{tabular}
\end{center}
\end{table}
\begin{table}[H]
\begin{center}
\caption{Number of sets rejected by the proportion test. Each set was generated by Mersenne twister. Boldface values mean out of the standard deviation range.}
\label{MT2}
\begin{tabular}{lccc}
\hline
Length &Kim& Pareschi & Proposed\\
\hline
\hline
$10^3$&\bf 185 &2 &\bf 31\\
$5\times10^3$&\bf 57&2&\bf 7\\
$10^4$&\bf 11&\bf  11&\bf  8\\
$5\times10^4$&\bf 37&  3& 3\\
$10^5$&\bf 25 &4& 2\\
$5\times10^5$&\bf 21&\bf  0&\bf  8\\
$10^6$&\bf 30&\bf  6& 3\\
$5\times10^6$&\bf 25& 5&\bf  1\\
$10^7$& \bf 23& 2&\bf  1\\
\hline
\end{tabular}
\end{center}
\end{table}
\begin{table}[H]
\begin{center}
\caption{Number of sets rejected by the proportion test. Each set was  generated by AES. Boldface values mean out of the standard deviation range.}
\label{AES2}
\begin{tabular}{lccc}
\hline
Length &Kim& Pareschi & Proposed\\
\hline
\hline
$10^3$&\bf 173 &\bf 1 &\bf 33\\
$5\times10^3$&\bf 59&4&\bf 6\\
$10^4$&\bf 10&\bf  10&4\\
$5\times10^4$&\bf 30&  5& \bf 6\\
$10^5$&\bf 32 &3& 2\\
$5\times10^5$&\bf 24&\bf  2&3\\
$10^6$&\bf 27&3&2 \\
$5\times10^6$&\bf 32& 2&3\\
$10^7$& \bf 22& 2&3\\
\hline
\end{tabular}
\end{center}
\end{table}
\begin{table}[H]
\begin{center}
\caption{Number of sets rejected by the uniformity test. Each set was generated by Mersenne twister. Boldface values mean out of the standard deviation range.}
\label{MT3}
\begin{tabular}{lccc}
\hline
Length &Kim& Pareschi & Proposed\\
\hline
\hline
$10^3$&\bf 1000 &\bf1000 &\bf 2\\
$5\times10^3$&\bf 1000&\bf1000&0\\
$10^4$&\bf 969&\bf  893&0\\
$5\times10^4$&\bf 8&  \bf236& 0\\
$10^5$&\bf 9 &\bf1& 0\\
$5\times10^5$&\bf 1&\bf  1&0\\
$10^6$&0& 0& 0\\
$5\times10^6$&0& 0& 0\\
$10^7$& \bf 1& 0&  0\\
\hline
\end{tabular}
\end{center}
\end{table}
\begin{table}[H]
\begin{center}
\caption{Number of sets rejected by the uniformity test. Each set was  generated by AES. Boldface values mean out of the standard deviation range.}
\label{AES3}
\begin{tabular}{lccc}
\hline
Length &Kim& Pareschi & Proposed\\
\hline
\hline
$10^3$&\bf 1000 &\bf1000 &\bf 0\\
$5\times10^3$&\bf 1000&\bf1000&0\\
$10^4$&\bf 987&\bf  916&0\\
$5\times10^4$&\bf 7&  \bf213& 0\\
$10^5$&\bf 5 &0& 0\\
$5\times10^5$&\bf 1&0& 0\\
$10^6$&\bf 1& 0& 0\\
$5\times10^6$&0&0 &0 \\
$10^7$& 0& 0&  0\\
\hline
\end{tabular}
\end{center}
\end{table}
\begin{table}[H]
\begin{center}
\caption{P-values generated by the  uniformity test for $p_{uniform}$. Here, $p_{uniform}$ is a p-value computed in the uniformity test for each set generated by Mersenne twister.  Boldface values mean rejection.}
\label{MT4}
\begin{tabular}{lccc}
\hline
Length &Kim& Pareschi & Proposed\\
\hline
\hline
$10^3$&\bf 0.000000 &\bf 0.000000&0.137282\\
$5\times10^3$&\bf 0.000000&\bf 0.000000&0.383827\\
$10^4$&\bf 0.000000&\bf 0.000000&0.370262\\
$5\times10^4$&\bf 0.000000&\bf 0.000000&0.077607\\
$10^5$&\bf 0.000000&\bf 0.000000& 0.463512\\
$5\times10^5$&\bf 0.000000&\bf 0.000000&0.735908\\
$10^6$&\bf 0.000000&0.001046& 0.757790\\
$5\times10^6$&\bf 0.000000& 0.731886& 0.015926\\
$10^7$&\bf 0.000000& 0.707513& 0.976878\\
\hline
\end{tabular}
\end{center}
\end{table}
\begin{table}[H]
\begin{center}
\caption{P-values generated by the uniformity test for $p_{uniform}$. Here, $p_{uniform}$ is a p-value computed in the uniformity test for each set generated by AES.  Boldface values mean rejection.}
\label{AES4}
\begin{tabular}{lccc}
\hline
Length &Kim& Pareschi & Proposed\\
\hline
\hline
$10^3$&\bf 0.000000 &\bf 0.000000&0.227180\\
$5\times10^3$&\bf 0.000000&\bf 0.000000&0.552383\\
$10^4$&\bf 0.000000&\bf 0.000000&0.012387\\
$5\times10^4$&\bf 0.000000&\bf 0.000000&0.228367\\
$10^5$&\bf 0.000000&\bf 0.000000& 0.146982\\
$5\times10^5$&\bf 0.000000&\bf 0.000000&0.029011\\
$10^6$&\bf 0.000000& 0.042255& 0.455937\\
$5\times10^6$&\bf 0.000000&0.964295 &0.997147\\
$10^7$&\bf 0.000000& 0.628790& 0.41005\\
\hline
\end{tabular}
\end{center}
\end{table}

\subsection{Experiment 2}
Firstly, we generated sequences with Mersenne twister.
After that, for each sequence, we did the following replacement:
\begin{align*}
x_i=&1 \ \ \ \ (i\equiv T\mod2T),\\
x_i=&-1 \ \ (i\equiv 0\mod2T),
\end{align*}
where $T$ is a parameter.
Clearly, the sequences do not have good randomness.
For each fixed $T$, we performed test 100 times.
One test was performed for 1000 sequences and each sequence is 1000000-bit.
For fair evaluation, the same sequences were used for evaluation of the proposed test, the present DFTT and Pareschi et al.'s test.

\begin{figure}[H]
\begin{center}
\includegraphics[width=10cm]{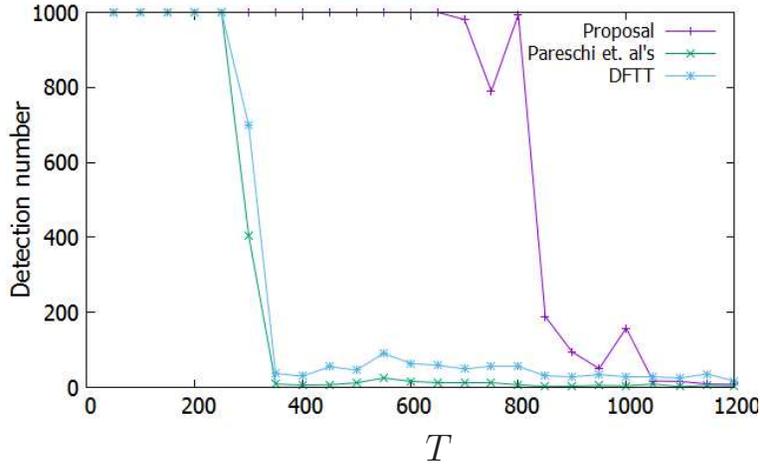}
\begin{picture}(0,0)
\put(-130,-5){\large $T$}
\end{picture}
\end{center}
\caption{Detection number by proportion.}
\label{D1}
\end{figure}

\begin{figure}[H]
\begin{center}
\includegraphics[width=10cm]{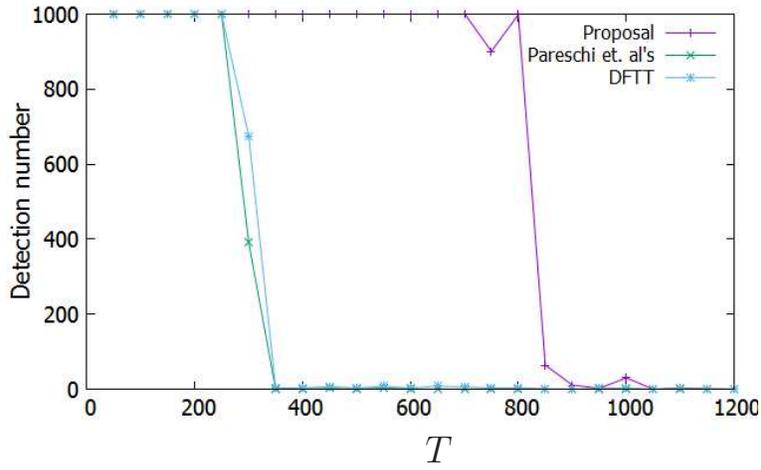}
\begin{picture}(0,0)
\put(-130,-5){\large $T$}
\end{picture}
\end{center}
\caption{Detection number by uniformity.}
\label{D2}
\end{figure}

\begin{figure}[H]
\begin{center}
\includegraphics[width=10cm]{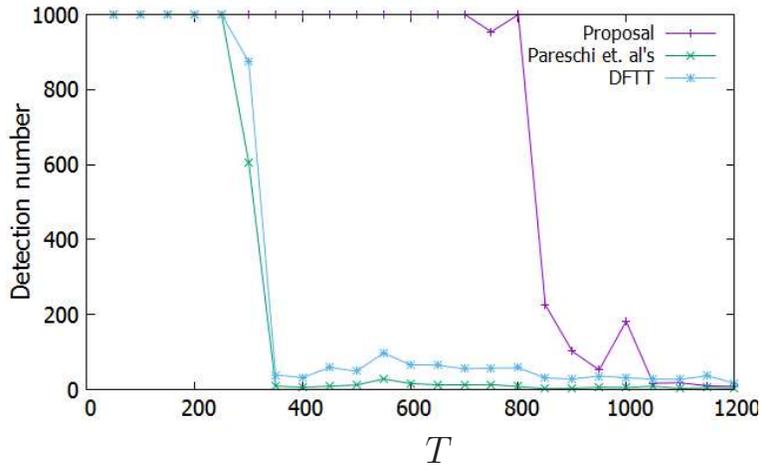}
\begin{picture}(0,0)
\put(-130,-5){\large $T$}
\end{picture}
\end{center}
\caption{Total detection number.}
\label{D3}
\end{figure}

As the result, we got Fig. \ref{D1}, \ref{D2} and \ref{D3}. 
The proposed test could detect non-negligible deviations for large $T$ as compared with the present DFTT and Pareschi et al.'s test.
It shows that detection power of the proposed test is better than those of the others.

\section{Conclusion}
The theoretical reference distribution of the present DFTT have not been derived so far although the reference distribution is fundamental for a statistical test.
Then, we proposed  a new test whose reference distribution can be theoretically derived.
The proposed test uses the fact that the moments of scaled variance of power spectrum converge to those of the standard normal distribution.
The proposed test needs $O(n\log n)$ times, that is the same as the present DFTT.
Some experiments showed that the proposal method has better detection power than the present DFTT.
From the above, the proposed test is useful for practical randomness evaluation and could be considered as an alternative of the DFTT with theoretical distribution.

As a future work, we have to investigate independency between the proposed test and the other tests included in NIST SP800-22 because randomness of given sequences should be judged by all the results of tests included in NIST SP800-22.

\end{document}